\def\year{2020}\relax
\documentclass[letterpaper]{article} 
\usepackage{aaai20}  
\usepackage{times}  
\usepackage{helvet} 
\usepackage{courier}  
\usepackage[hyphens]{url}  
\usepackage{graphicx} 
\usepackage{graphicx}  
\usepackage{algorithmicx,algorithm}
\usepackage{amsmath,amssymb,amsfonts}

\def\ie{\emph{i.e.}}
\def\eg{\emph{e.g.}}

\def\aka{{\em aka.}}

\usepackage{booktabs}
\usepackage{bm}
\usepackage{array}
\usepackage[noend]{algpseudocode}
\usepackage{algorithmicx,algorithm}
\usepackage{multirow}

\title{Towards Cross-modality Medical Image Segmentation with Online Mutual Knowledge Distillation}

\author{Kang Li\textsuperscript{\rm 1}, Lequan Yu\textsuperscript{\rm 1}\thanks{Corresponding Author.}, Shujun Wang\textsuperscript{\rm 1}, Pheng-Ann Heng\textsuperscript{\rm 1,2}\\ 
\textsuperscript{\rm 1} Department of Computer Science and Engineering, The Chinese University of Hong Kong\\ 
\textsuperscript{\rm 2} Guangdong Provincial Key Laboratory of Computer Vision and Virtual Reality Technology,\\ Shenzhen Institutes of Advanced Technology, Chinese Academy of Sciences, China\\
\{kli, lqyu, sjwang, pheng\}@cse.cuhk.edu.hk
}

\begin{document}

\maketitle
\begin{abstract}
The success of deep convolutional neural networks is partially attributed to the massive amount of annotated training data.
However, in practice, medical data annotations are usually expensive and time-consuming to be obtained.
Considering multi-modality data with the same anatomic structures are widely available in clinic routine,
in this paper, we aim to exploit the prior knowledge (\eg, shape priors) learned from one modality (\aka, assistant modality) to improve the segmentation performance on another modality (\aka, target modality) to make up annotation scarcity.
To alleviate the learning difficulties caused by modality-specific appearance discrepancy, we first present an Image Alignment Module (IAM) to narrow the appearance gap between assistant and target modality data.
We then propose a novel Mutual Knowledge Distillation (MKD) scheme to thoroughly exploit the modality-shared knowledge to facilitate the target-modality segmentation.
To be specific, we formulate our framework as an integration of two individual segmentors.
Each segmentor not only explicitly extracts one modality knowledge from corresponding annotations, but also implicitly explores another modality knowledge from its counterpart in mutual-guided manner.
The ensemble of two segmentors would further integrate the knowledge from both modalities and generate reliable segmentation results on target modality.
Experimental results on the public multi-class cardiac segmentation data, \ie, MM-WHS 2017, show that our method achieves large improvements on CT segmentation by utilizing additional MRI data and outperforms other state-of-the-art multi-modality learning methods.
\end{abstract}
\section{Introduction}
Modern clinical practice usually involves multiple imaging techniques, \eg, magnetic resonance imaging (MRI) and computed tomography (CT), to acquire a more comprehensive view of particular tissues or organs in disease diagnosis and surgical planning~\cite{cao2017dual}. For instance, CT and MRI are extensively used to provide clear anatomical information of cardiac structures~\cite{zhuang2019evaluation}.
As different imaging modalities are based on different physical imaging principles, they normally emphasize organs or tissues with distinct visual contrast. For example, in MRI brain scans, MRI-T2 has better visual effects in edema, while MRI-T1c highlights gross tumor core~\cite{woo2014multimodal}.
Therefore, multi-modality learning is progressively developed in medical imaging domain.
In recent years, deep learning based methods have achieved promising performance in many medical image analysis tasks, relying on the massive amount of annotated data~\cite{litjens2017survey}. However, large annotated datasets are usually expensive and time-consuming to acquire in medical domain, since the annotations have to be marked by professional medical experts under strict and meticulous inspection.
Considering different imaging modalities reflect the same anatomic structures, multi-modality learning has became a promising direction to make up the annotation scarcity in automatic medical image analysis.
In this paper, we aim to investigate the effectiveness of \textit{utilizing the prior knowledge (e.g., shape priors) learned from one modality (aka. assistant modality) to improve the segmentation performance on another modality (aka. target modality)}, where no assistant-modality data is required in testing phase.

In general, one intuitive way is joint-training, where the network is trained with both assistant-modality and target-modality data.
Another straightforward approach is to fine-tune a deep convolutional neural network (DCNN) learned from assistant-modality data with the target-modality data, so that it can transfer the prior knowledge learned from assistant modality to target-modality tasks.
Nevertheless, the shared cross-modality information can not be well exploited in either joint-training or fine-tuning, since the representative modality-shared features are hard to directly learn from multi-modality data with large appearance discrepancy.
~\citeauthor{valindria2018multi}~\shortcite{valindria2018multi} proposed to assign each modality data with its specific feature extractor to alleviate the negative influences from distinct multi-modality appearances.
They presented four dual-stream architectures with well-designed parameter sharing strategies to explore cross-modality information, and demonstrated X-shape architecture is more suitable for multi-modality image segmentation.
However, these dual-stream architectures incorporate human intervention in determining parameter sharing, and would require special adjustments to generalize to other medical analysis tasks.
In our problem setting, two modality information are involved in the network learning, where the valuable modality-shared knowledge should be thoroughly explored to enhance the generalization of networks, while the redundant modality-specific appearances should be aligned to ease network learning.

In this paper, we propose a novel cross-modality image segmentation framework based on knowledge distillation to utilize assistant-modality (\eg, MRI) priors to improve the segmentation performance on target-modality data (\eg, CT).
To alleviate the difficulties caused by modality-specific appearance discrepancy and facilitate the learning of modality-shared knowledge, we first present an Image Alignment Module (IAM) to reduce the appearance gap by translating the assistant-modality images to synthetic target-modality images via adversarial learning~\cite{goodfellow2014generative}.
As the core workhorses in our framework, we develop a novel Mutual Knowledge Distillation (MKD) scheme to better exploit the modality-shared knowledge from both assistant and target modality data for superior segmentation results.
We formulate this scheme as the cohort of two individual segmentors, which directly learn feature representations from synthetic target-modality data and real target-modality data, respectively.
Besides explicitly learned from segmentation annotations, two segmentors are also mutually guided by its counterpart outputs implicitly via knowledge distillation~\cite{hinton2015distilling}.
Armed with the mutual guidance, both segmentors are able to explicitly learn the knowledge from one modality and implicitly learn the knowledge from another modality concurrently.
Additionally, the ensemble of two segmentors would further integrate the knowledge learned from two modalities and perform more reliable segmentation in target modality.
The whole framework is optimized in end-to-end online manner, so that the IAM could receive timely feedback from MKD scheme to generate more reasonable synthetic target-modality images with well-preserved shape priors.
We extensively evaluate our method on the MM-WHS 2017 Challenge dataset~\cite{zhuang2019evaluation}. Our framework brings large dice improvements ($3.06\%$) for cardiac segmentation on CT volumes by utilizing additional MRI data, substantially outperforming other multi-modality learning methods.
The major contributions of this paper are as follows.
\begin{itemize}
    \item We present an effective multi-modality learning framework to utilize prior knowledge of one modality to enhance the segmentation performance of another modality.

    \item We propose a novel Mutual Knowledge Distillation scheme to better exploit modality-shared knowledge with mutual guidance of segmentor outputs.

    \item We also present an Image Alignment Module to reduce cross-modality appearance discrepancy to promote the learning of modality-shared knowledge.

    \item We extensively evaluate our method on public multi-class cardiac segmentation challenge data~\cite{zhuang2019evaluation}. Our method outperforms other state-of-the-art multi-modality learning methods.

\end{itemize}

\section{Related Work}

\begin{figure*}[t]
\centering
\includegraphics[width=0.93\textwidth]{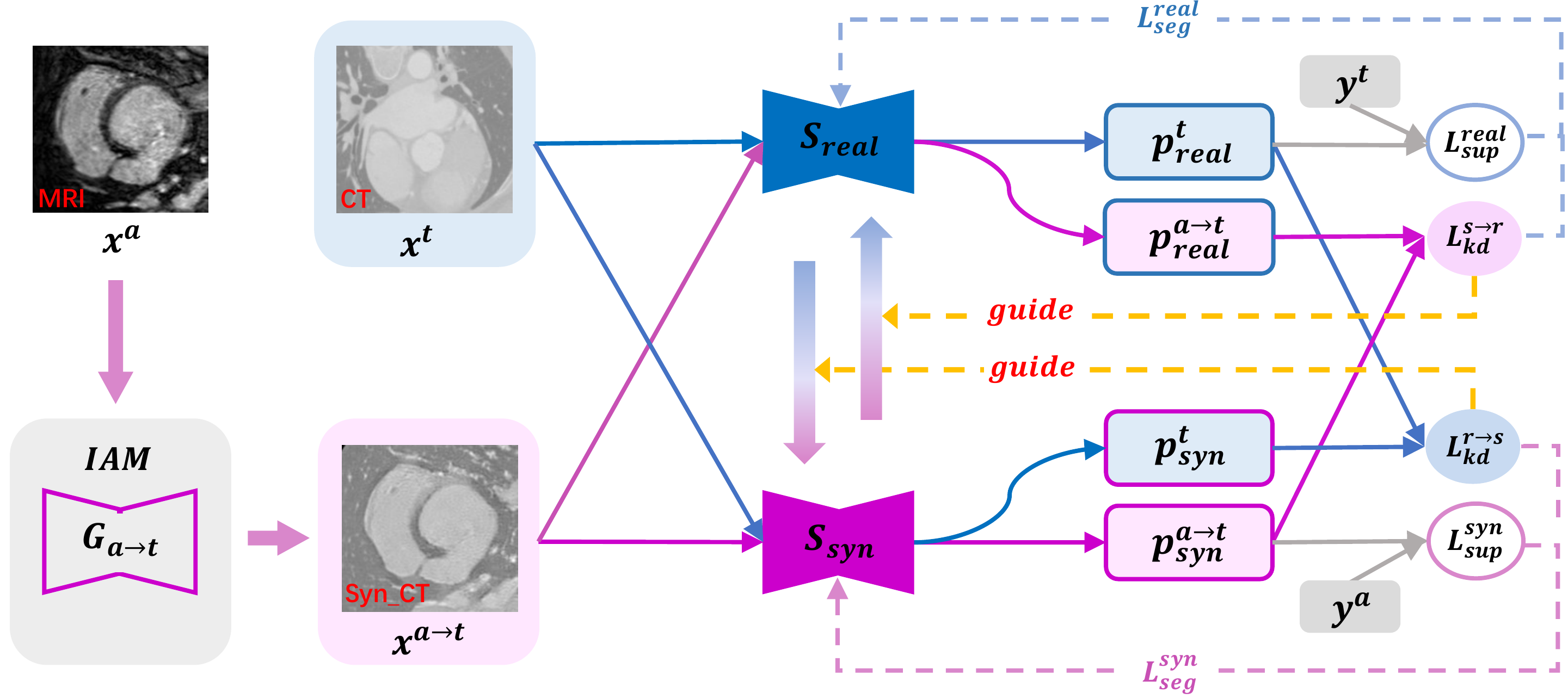}
\caption{Overview of our framework, where magenta and blue represent the data flow of assistant modality (\eg, MRI) and target modality (\eg, CT) respectively. The generator $G_{a \rightarrow t}$ performs assistant-to-target translation and outputs synthesized CT data. The magenta-to-blue transition arrow represents the knowledge transfer from $S_{syn}$ to $S_{real}$ guided by $L_{kd}^{s \rightarrow r}$, while the blue-to-magenta transition arrow demonstrates the knowledge transfer from $S_{real}$ to $S_{syn}$ guided by $L_{kd}^{r \rightarrow s}$. In this way, both segmentors are mutually guided. In testing, we feed $x_{t}$ into both segmentors and adopt the ensemble prediction as final results.
}
\label{fig:framework}
\end{figure*}

\subsection{Multi-modality Learning in Medical Imaging}
Multi-modality learning has been widely studied in medical field.
Many approaches~\cite{fidon2017scalable,kamnitsas2017efficient,guo2018medical,nie2016fully} proposed different feature fusion strategies to utilize complementary visual reflections of multi-modality data for comprehensive segmentation.
For example, \citeauthor{kamnitsas2017efficient}~\shortcite{kamnitsas2017efficient} presented a dual pathway architecture to incorporate cross-modality local and global features for brain lesion segmentation from multi-channel MRI data, while \citeauthor{fidon2017scalable}~\shortcite{fidon2017scalable} designed a nested architecture to jointly utilize different modality of MRI brain tumor data and adapt their framework to handle scalable inputs.
\citeauthor{guo2018medical}~\shortcite{guo2018medical} compared the effectiveness of feature-level, classifier-level and decision-level feature fusion for multi-modality tumor segmentation.
However, these methods also require paired multi-modality data in testing stage to perform thorough feature fusion, while we aim to exploit the priors of assisted modality to to promote the performance on another modality by enhancing model generalization ability, where only target-modality data is required in the testing phase.

To utilize the prior knowledge between different modalities, \citeauthor{zheng2015cross}~\shortcite{zheng2015cross} proposed the marginal space learning method to transfer assisted modality knowledge to target modality.
\citeauthor{moeskops2016deep}~\shortcite{moeskops2016deep} investigated how to utilize multi-modality information under multi-task learning frameworks.
Recently, \citeauthor{valindria2018multi}~\shortcite{valindria2018multi} developed dual-stream encoder-decoder architectures, which assign each modality with specific branch and extract cross-modality features with delicately designed parameter sharing strategies.
With the recent advance in image translation~\cite{goodfellow2014generative,zhu2017unpaired}, many researchers were inspired to tackle medical problems with modality translation. Zhao et al.~\cite{zhao2019data} integrated registration between different modality MR images for data augmentation.
\citeauthor{jiang2018tumor}~\shortcite{jiang2018tumor} and Zhang et al.~\shortcite{zhang2018translating} proposed to eliminate the appearance gap of different modalities with generative adversarial network (GAN) for multi-modal segmentation.
\citeauthor{jiang2018tumor}~\shortcite{jiang2018tumor} proposed to train a GAN to generate meaningful synthetic MRI images first, and take synthetic images with real ones together as segmentor inputs for lung cancer segmentation, while Zhang et al.~\shortcite{zhang2018translating} proposed to jointly optimize GAN and segmentor, where the updated synthetic information could be exploited.
Different from these methods which directly fuse multi-modality knowledge by joint training, our framework takes two branches for synthetic data and real data respectively, and integrates cross-modality knowledge in a mutual-guided way.

\subsection{Knowledge Distillation}
Our work is also related to knowledge distillation, which was first proposed for network compression~\cite{hinton2015distilling}.
Hinton et al.~\shortcite{hinton2015distilling} proposed to use the output class probabilities of a static cumbersome model as soft targets to teach a lightweight student model.
~\citeauthor{Zagoruyko2017AT}~\shortcite{Zagoruyko2017AT} further forced the student network to mimic middle-level attention maps of the teacher network besides its output probability maps.
Since classical knowledge distillation approaches always require a powerful pre-trained teacher model, mutual learning was proposed to encourage an ensemble of networks to learn from each other cooperatively throughout the training process~\cite{zhang2018deep}.
Based on this, \citeauthor{wusi2019mutual}~\shortcite{wusi2019mutual} utilized mutual learning to capture complementary features in semi-supervised classification.
\citeauthor{wurm2019mutual}~\shortcite{wurm2019mutual} applied mutual learning between contour extraction and edge extraction for saliency detection.
We share a similar philosophy with mutual learning, but aim to design a cross-modality learning framework to utilize the shape priors from different modalities, which results in a different methodology.

\section{Methodology}
In this section, we introduce our proposed cross-modality image segmentation framework in details.
Given a set of labeled samples $\left\{x_{i}^{a}, y_{i}^{a}\right\}_{i=1}^{N}$ from assistant-modality data $X^{a}$, and a set of labeled samples $\left\{x_{j}^{t}, y_{j}^{t}\right\}_{j=1}^{M}$ from target-modality data $X^{t}$, we involve both $X^{a}$ and $X^{t}$ in network training, to improve the segmentation performance on target modality during testing.
As illustrated in Figure~\ref{fig:framework}, we design an Image Alignment Module (IAM) to eliminate the modality-specific appearance differences, and a Mutual Knowledge Distillation (MKD) scheme to thoroughly exploit modality-shared knowledge.
The whole framework is optimized in online mutual learning manner.

\subsection{Image Alignment Module}
Since redundant modality-specific appearances would introduce bias and increase the difficulty in leveraging the valuable modality-shared prior knowledge from $X^{a}$, we propose to perform a transformation from assistant-modality image $x^{a}$ to target-modality image $x^{t}$. In this way, the synthetic target-modality image $x^{a \rightarrow t}$ would acquire similar appearances to target-modality data with unaffected assistant-modality structures.

Inspired by generative adversarial networks~\cite{goodfellow2014generative}, we adopt a generator $G_{a \rightarrow t}$ and discriminator $D_{t}$ to perform the assistant-to-target image translation $G_{a \rightarrow t}: X^{a} \rightarrow X^{t}$ by adversarial learning.
To be specific, the generator $G_{a \rightarrow t}$ takes real assistant-modality image $x^{a}$ as inputs and produces realistic synthetic target-modality image $x^{a \rightarrow t} = G_{a \rightarrow t}(x^{a})$ to fool the discriminator $D_{t}$, while the discriminator $D_{t}$ tries to distinguish the synthetic images from real ones.
In this way, the $G_{a \rightarrow t}$ and the $D_{t}$ are competing while improving each other.
The parameters of $G_{a \rightarrow t}$ and $D_{t}$ are optimized by adversarial loss as
\begin{equation}
\begin{aligned}
\mathcal{L}_{\text {adv}}^{t}(G_{a \rightarrow t}, D_{t})=& \mathbb{E}_{x^{t} \sim X^{t}}[\log D_{t}(x^{t})]+\\ & \mathbb{E}_{x^{a} \sim X^{a}}[\log (1-D_{t}(G_{a \rightarrow t}(x^{a})))],
\end{aligned}
\label{eq:adv_a-t}
\end{equation}
where $G_{a \rightarrow t}$ tries to minimize the objective function by generating more realistic target-modality images, while $D_{t}$ intends to maximize it by identifying synthetic target-modality images from real ones.

Since the above mapping alone is weakly-constrained, an inverse mapping $G_{t \rightarrow a}: X^{t} \rightarrow X^{a}$ is introduced to impose the cycle-consistency constrain~\cite{zhu2017unpaired}.
Particularly, we attach another generator $G_{t \rightarrow a}$ and discriminator $D_{a}$ into this module and form a similar minimax game by adversarial loss in the following
\begin{equation}
\begin{aligned}
\mathcal{L}_{\text {adv}}^{a}(G_{t \rightarrow a}, D_{a})=& \mathbb{E}_{x^{a} \sim X^{a}}[\log D_{a}(x^{a})]+\\ & \mathbb{E}_{x^{t} \sim X^{t}}[\log (1-D_{a}(G_{t \rightarrow a}(x^{t})))].
\end{aligned}
\label{eq:adv_t-a}
\end{equation}
The cycle-consistency constraint forces the synthetic images could be translated back to the input images, meaning the reconstructed assistant-modality image $\hat{x}^{a} = G_{t \rightarrow a}\left(G_{a \rightarrow t}\left(x^{a}\right)\right)$ should be similar to $x^{a}$. And the reconstructed target-modality image $\hat{x}^{t} = G_{a \rightarrow t}\left(G_{t \rightarrow a}\left(x^{t}\right)\right)$ should also be closer to $x^{t}$. Specifically, the cycle-consistency loss is defined as
\begin{equation}
\begin{aligned}
    \mathcal{L}_{cyc}\left(G_{a \rightarrow t}, G_{t \rightarrow a}\right)=&\mathbb{E}_{x^{a} \sim X^{a}}\left\| \hat{x}^{a}-x^{a}\right \|_{1} + \\ &\mathbb{E}_{x^{t} \sim X^{t}}\left\| \hat{x}^{t}-x^{t}\right \|_{1}.
\end{aligned}
\label{eq:cyc}
\end{equation}
Overall, the full objective functions to optimize $G_{a \rightarrow t}$ and $G_{t \rightarrow a}$ are summarized as
\begin{align}
    \mathcal{L}_{gan}^{a \rightarrow t} &= \mathcal{L}_{adv}^{t} + \lambda_{cyc}\mathcal{L}_{cyc},\label{eq:gan-a-t}\\
    \mathcal{L}_{gan}^{t \rightarrow a} &= \mathcal{L}_{adv}^{a} + \lambda_{cyc}\mathcal{L}_{cyc},
    \label{eq:gan-t-a}
\end{align}
where $\lambda_{cyc}$ is a hyperparameter to balance training. Noted that the generator $G_{a \rightarrow t}$ is co-optimized with the following Mutual Knowledge Distillation scheme to preserve shape invariance in end-to-end manner, which will be concretely described in Section~\ref{sec:onlinelearning}.
Through this module, the appearances of assistant-modality data are aligned with target modality, and the modality-shared prior knowledge could be better investigated in the following step.

\subsection{Mutual Knowledge Distillation}
To thoroughly exploit modality-shared knowledge, we develop a novel Mutual Knowledge Distillation (MKD) scheme to facilitate the network learning.
We formulate it as a cohort of two segmentation networks, \ie, synthetic segmentor $S_{syn}$ and real segmentor $S_{real}$, where  $S_{syn}$ and $S_{real}$ receive direct supervision from synthetic target-modality image $x^{a \rightarrow t}$ and real target-modality image $x^{t}$, respectively.
During training, we not only encourage each segmentor to explicitly learn one modality knowledge from corresponding annotations, but also guide it to implicitly explore another modality knowledge from its peer segmentor outputs.
With mutual guidance, both segmentors could explore feature representations of two modalities in explicit and implicit ways.

For explicit learning, the segmentors are optimized with the supervised segmentation loss with respect to the ground truth.
Specifically, the synthetic segmentor $S_{syn}$ mainly learns assistant-modality features from synthetic target-modality image $x^{a \rightarrow t}$ with assistant-modality annotation $y^{a}$, while the real segmentor $S_{real}$ mainly exploits target-modality information from real target-modality image $x^{t}$ with target-modality ground truth $y^{t}$.
The supervised segmentation loss is formulated as the combination of cross-entropy loss $\mathcal{L}_{\text {ce}}$ and dice loss $\mathcal{L}_{\text {dice}}$:
\begin{equation}
\begin{aligned}
p^{a \rightarrow t}_{syn} &=S_{syn}\left(x^{a \rightarrow t}\right),\\
p^{t}_{real} &=S_{real}\left(x^{t}\right),\\
\mathcal{L}_{\text {sup}}^{\text{syn}}&=\mathcal{L}_{\text {ce}}\left(y^{a}, p^{a \rightarrow t}_{syn}\right)+\mathcal{L}_{\text {dice}}\left(y^{a},  p^{a \rightarrow t}_{syn}\right),\\
\mathcal{L}_{\text {sup}}^{\text{real}}&=\mathcal{L}_{\text {ce}}\left(y^{t}, p^{t}_{real}\right)+\mathcal{L}_{\text {dice}}\left(y^{t},  p^{t}_{real}\right),
\end{aligned}
\label{eq:sup}
\end{equation}
where $y^{a}$ and $y^{t}$ are the segmentation ground truth of assistant modality and target modality, respectively.

For implicit learning, we propose to utilize the outputs of each segmentor as guidance for its counterpart, so that synthetic segmentor $S_{syn}$ and real segmentor $S_{real}$ are mutually guided by each other and interactively explored the dark knowledge (\ie, output category similarity) from each other~\cite{furlanello2018born}~\cite{hinton2015distilling}.
To make real segmentor exploit the guidance from synthetic segmentor, we feed synthetic target-modality image $x^{a \rightarrow t}$ into real segmentor $S_{real}$ to produce the probability map $p_{real}^{a \rightarrow t} = S_{real}\left(x^{a \rightarrow t}\right)$, and encourage it to be similar to synthetic segmentor output $p_{syn}^{a \rightarrow t}$ with the synthetic-to-real knowledge distillation loss $\mathcal{L}_{\text {kd}}^{s \rightarrow r}$ (\ie, the magenta filled circle in Figure~\ref{fig:framework}).
Since $S_{syn}$ is directly supervised by synthetic target-modality annotations, the guidance (\ie, $p_{syn}^{a \rightarrow t}$) is trustworthy.
Furthermore, as $S_{syn}$ is equipped with assistant-modality priors, this synthetic-to-real knowledge distillation scheme would transfer knowledge from $S_{syn}$ to $S_{real}$ and facilitate $S_{real}$ to integrate the cross-modality information effectively.
Similarly, we also feed real target-modality images $x^{t}$ into synthetic segmentor $S_{syn}$ to obtain the probability map $p_{syn}^{t} = S_{syn}\left(x^{t}\right)$.
By narrowing down the differences between $p_{syn}^{t}$ and real segmentor output $p_{real}^{t}$, $S_{syn}$ takes the informative guidance from $S_{real}$ by real-to-synthetic knowledge distillation loss $\mathcal{L}_{\text {kd}}^{r \rightarrow s}$ (\ie, the blue filled circle in Figure~\ref{fig:framework}).
With the training going on, both segmentors are mutually guided to gradually explore the knowledge of another modality from its counterpart.
We formulate the knowledge distillation loss as the cross-entropy between two probability maps, following previous work~\cite{hinton2015distilling}:
\begin{equation}
\begin{aligned}
\mathcal{L}_{\text {kd}}^{s \rightarrow r}&=\mathcal{L}_{\text {ce}}\left(
p_{syn}^{a \rightarrow t}, p_{real}^{a \rightarrow t}\right),\\
\mathcal{L}_{\text {kd}}^{r \rightarrow s}&=\mathcal{L}_{\text {ce}}\left(p_{real}^{t}, p_{syn}^{t}\right).
\end{aligned}
\label{eq:kd}
\end{equation}

The objective functions to train $S_{syn}$ and $S_{real}$ are then defined as
\begin{equation}
\begin{aligned}
\mathcal{L}_{\text {seg}}^{real} &= \mathcal{L}_{\text {sup}}^{real} + \lambda_{kd}^{1} \mathcal{L}_{\text {kd}}^{s \rightarrow r},\\
\mathcal{L}_{\text {seg}}^{syn} &= \mathcal{L}_{\text {sup}}^{syn} + \lambda_{kd}^{2}\mathcal{L}_{\text {kd}}^{r \rightarrow s},
\end{aligned}
\label{eq:seg_total}
\end{equation}
where $\lambda_{kd}^{1}$ and $\lambda_{kd}^{2}$ are hyperparameters.
In summary, each individual segmentor is optimized by the supervised segmentation loss to explicitly learn one modality knowledge, as well as knowledge distillation loss to implicitly absorb another modality knowledge, and thus comprehensively investigates cross-modality information under peer guidance.

\subsection{Online Mutual Training}
\label{sec:onlinelearning}
Overall, our total objective to train the whole framework can be formulated as
\begin{equation}
    \mathcal{L}_{obj} = \mathcal{L}_{gan}^{a \rightarrow t} + \mathcal{L}_{gan}^{t \rightarrow a} + \mathcal{L}_{seg}^{syn} + \mathcal{L}_{seg}^{real}.
    \label{eq:total}
\end{equation}
The entire framework is trained in online manner, and we update all components alternatively in each iteration.
Algorithm \ref{Al:1} presents detailed training procedures. We closely follow the settings in~\citeauthor{zhu2017unpaired}~\shortcite{zhu2017unpaired} to optimize $G_{a \rightarrow t}$, $D_{t}$, $G_{t \rightarrow a}$, and $D_{a}$, except that the generator $G_{a \rightarrow t}$ is optimized together with $S_{syn}$ by $\mathcal{L}_{gan}^{a \rightarrow t}$ and $\mathcal{L}_{\text {sup}}^{syn}$ with $S_{syn}$ fixed.
This strategy ensures geometric transformation consistency in $G_{a \rightarrow t}$ by constraining pixel-level semantic relationship of synthetic target-modality images.
Due to that, $G_{a \rightarrow t}$ could generate meaningful synthetic target-modality data with unaffected shape priors.

For segmentation networks, we alternatively update the weights of $S_{syn}$ and $S_{real}$ by the combination of supervised segmentation loss and knowledge distillation loss.
From $S_{syn}$ view, the real-to-synthetic knowledge distillation loss provides chances to have a glimpse of real data, which guides $S_{syn}$ to generalize towards a more reliable direction.
From $S_{real}$ view, the synthetic-to-real knowledge distillation loss brings additional knowledge of highly realistic synthetic data as data augmentation and directly enhance its generalization ability. The real-to-synthetic and synthetic-to-real knowledge distillation make $S_{syn}$ and $S_{real}$ mutually benefit from each other's guidance in online manner.
The ensemble model is able to explore more informative and comprehensive cross-modality knowledge. In testing, we feed $x_{t}$ into both segmentors and adopt the ensemble prediction as final results.
\begin{algorithm}[t]
\caption{Training procedure of the proposed approach}
\label{Al:1}
{\bf Input:}
A batch of $\left(x^{t}, y^{t}\right)$ from target-modality dataset $X^{t}$ and $\left(x^{a}, y^{a}\right)$ from assistant-modality dataset $X^{a}$\\
{\bf Output:}
The prediction $p^{t}$ of input $x^{t}$
\begin{algorithmic}[1]
\State $\theta_{G_{a \rightarrow t}}, \theta_{G_{t \rightarrow a}}, \theta_{D_{a}}, \theta_{D_{t}}, \theta_{S_{syn}}, \theta_{S_{real}}$ $\leftarrow$ initialize

\While{not converge}
    \State $(x^{t}, y^{t})$, $(x^{a}, y^{a})$ $\leftarrow$ sampled from $X^{t}$ and $X^{a}$
    \State Calculate $\mathcal{L}_{gan}^{a \rightarrow t}$ and $\mathcal{L}_{sup}^{syn}$ as Eq.~\eqref{eq:gan-a-t} and~\eqref{eq:sup}
    \State Update $\theta_{G_{a \rightarrow t}}$ $\stackrel{+}{\leftarrow}-\Delta_{\theta_{G_{a \rightarrow t}}}\left( \mathcal{L}_{gan}^{a \rightarrow t} +\mathcal{L}_{sup}^{syn}\right)$

    \State Calculate $\mathcal{L}_{adv}^{t}$ as Eq.~\eqref{eq:adv_a-t}
    \State Update $\theta_{D_{t}}$ $\stackrel{+}{\leftarrow}-\Delta_{\theta_{D_{t}}}\mathcal{L}_{adv}^{t}$
    \State Calculate $\mathcal{L}_{gan}^{t \rightarrow a}$ as Eq.~\eqref{eq:gan-t-a}
\State Update $\theta_{G_{t \rightarrow a}}$ $\stackrel{+}{\leftarrow}-\Delta_{\theta_{G_{t \rightarrow a}}}\mathcal{L}_{gan}^{t \rightarrow a}$

\State Calculate $\mathcal{L}_{adv}^{a}$ as Eq.~\eqref{eq:adv_t-a}

\State Update $\theta_{D_{a}}$ $\stackrel{+}{\leftarrow}-\Delta_{\theta_{D_{a}}}\mathcal{L}_{adv}^{a}$

\State Calculate $\mathcal{L}_{seg}^{syn}$ and $\mathcal{L}_{seg}^{real}$ as Eq.~\eqref{eq:seg_total}

\State Update $\theta_{S_{syn}}$ $\stackrel{+}{\leftarrow}-\Delta_{\theta_{S_{syn}}}\mathcal{L}_{seg}^{syn}$

\State Update $\theta_{S_{real}}$ $\stackrel{+}{\leftarrow}-\Delta_{\theta_{S_{real}}}\mathcal{L}_{seg}^{real}$

\EndWhile

\State Calculate $p^{t}_{real}$ and $p^{t}_{syn}$

\State \Return $p^{t}$ $\leftarrow$ the ensemble of $p^{t}_{real}$ and $p^{t}_{syn}$
\end{algorithmic}
\end{algorithm}

\section{Experiments}

\begin{table*}[t]
\centering
\caption{Quantitative comparison between our method and other multi-modality segmentation methods. Here we take CT as target modality and MR as assistant modality. The dice of all heart substructures and the average of them are reported here.}
\smallskip
\label{tab:quatitative}
\begin{tabular}{c|c|c|c|c|c|c|c|c}
\toprule[1pt]
\multirow{2}{*}{Method} & \multirow{2}{*}{Mean Dice} & \multicolumn{7}{c}{Dice of substructures of heart} \\ \cline{3-9} 
 & & MYO & LA & LV & RA & RV & AA & PA\\
\midrule
Baseline & $0.8706$ & $0.8702$ & $0.8922$ & $0.9086$ & $0.8386$ & $0.846$ & $0.9252$ & $0.8134$ \\
Fine-tune & $0.8769$ & $0.8716$ & $0.9040$ & $0.9079$ & $0.8443$ & $0.8526$ & $0.9274$ & $0.8305$ \\
Joint-training & $0.8743$ & $0.8665$ & $0.9076$ & $0.9123$ & $0.8278$ & $0.8492$ & $0.9302$ & $0.8266$ \\
X-shape ~\shortcite{valindria2018multi} & $0.8767$ & $0.8719$ & $0.8979$ & $0.9094$ & $0.8551$ & $0.8444$ & $0.9343$ & $0.8240$ \\
Jiang et al.~\shortcite{jiang2018tumor} & $0.8765$ & $0.8723$ & $0.9054$ & $0.9073$ & $0.8338$ & $0.8525$ & $0.9484$ & $0.8156$ \\
Zhang et al.~\shortcite{zhang2018translating} & $0.8850$ & $0.8781$ & $0.9112$ & $0.9134$ & $0.8514$ & $0.8631$ & $0.9430$ & $0.8342$ \\
\midrule
Ours ($S_{syn}$) & $0.8849$ & $0.8738$ & $0.9086$ & $0.9169$ & $0.8571$ & $0.8633$ & $0.9412$ & $0.8333$ \\
Ours ($S_{real}$) & $0.8947$ & $0.8893$ & $0.9131$ & $0.9197$ & $0.8668$ & $0.8758$ & $0.9552$ & $0.8432$ \\
Ours & $\bm{0.9012}$ & $\bm{0.8934}$ & $\bm{0.9190}$ & $\bm{0.9267}$ & $\bm{0.8747}$ & $\bm{0.8814}$ & $\bm{0.9595}$ & $\bm{0.8538}$ \\
\bottomrule[1pt]
\end{tabular}
\end{table*}

\begin{figure*}[h]
\centering
\includegraphics[width=0.90\textwidth]{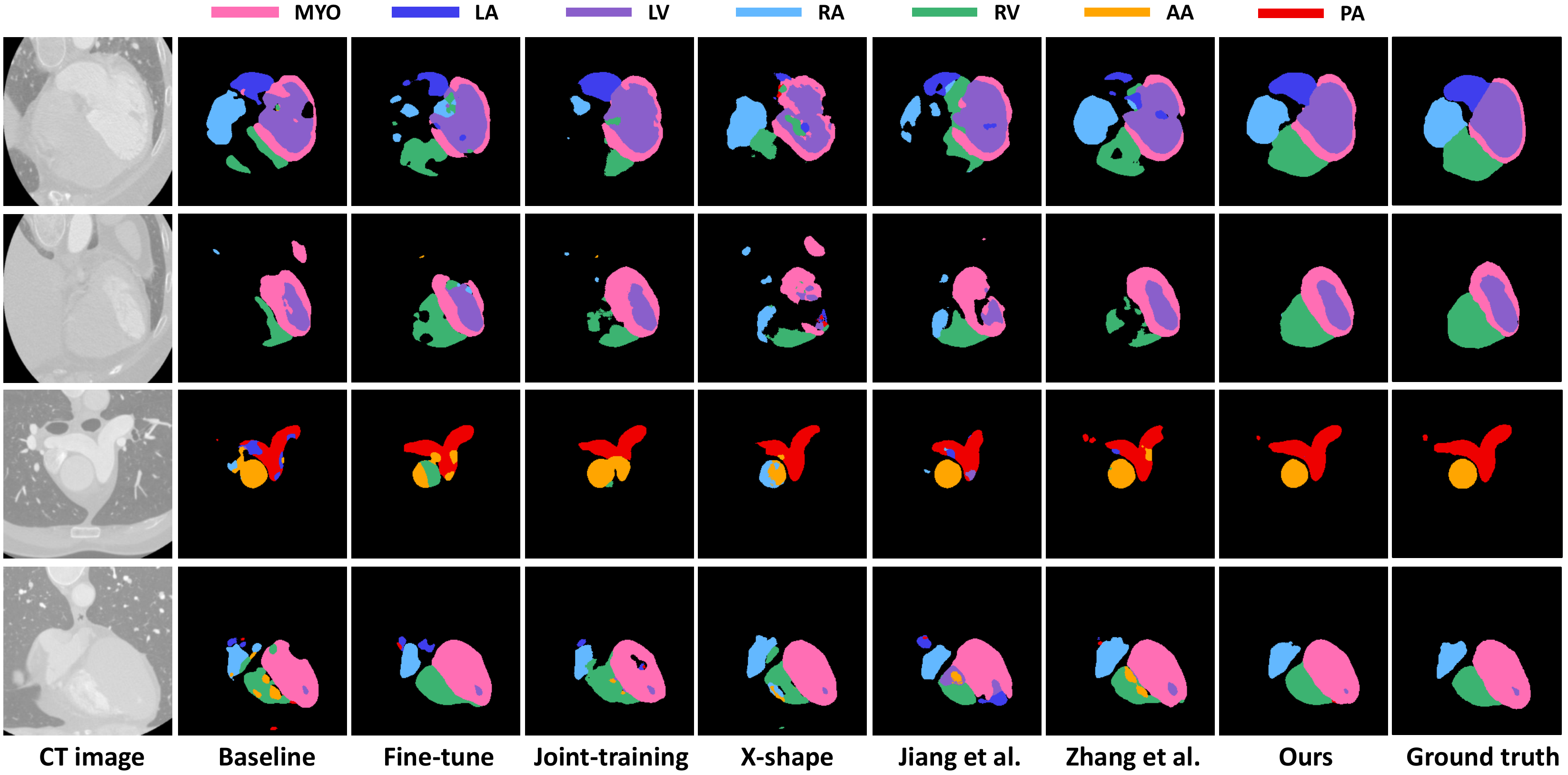}
\caption{Visual comparison of segmentation results produced by different methods, where the legend for cardiac substructures is presented on the top. As we can see, our results (second last column) are closer to the ground truth (last column) than others.
}
\label{fig:qualitative}
\end{figure*}

\subsection{Dataset and Implementation Details}
We evaluate the proposed method on the Multi-modality Whole Heart Segmentation Challenge 2017 (MM-WHS 2017) dataset, which contains unpaired 20 MRI and 20 CT volumes as the training data and the annotations of 7 cardiac substructures including the left ventricle blood cavity (LV), the right ventricle blood cavity (RV), the left atrium blood cavity (LA), the right atrium blood cavity (RA), the myocardium of the left ventricle (MYO), the ascending aeorta (AA), and the pulmonary artery (PA)~\cite{zhuang2019evaluation}.
Here, we take MRI as assistant modality and CT as target modality, as MRI images have better soft tissue contrast and would provide more detailed heart information for CT segmentation.
We randomly split 20 CT volumes into two folds and perform two-fold cross validation. In each fold, we use all 20 MRI volumes and 10 CT volumes to train our network.

Following data preprocessing procedures of previous works in this dataset~\cite{chen2019synergistic}, the image slices are sampled in transverse view, cropped centered at heart region, and resized into $256 \times 256$. Data augmentation methods applied here include random flipping and rotating.
The architectures of generators and discriminators are closely following the setting in~\citeauthor{zhu2017unpaired}~\shortcite{zhu2017unpaired},
and the segmentors adopt the baseline network architecture in~\cite{valindria2018multi}, consisting of Unet~\shortcite{ronneberger2015unet} with residual blocks~\shortcite{he2016identity}.
In training, Adam optimizer is used for the optimization of generators, discriminators and segmentors with the learning rate of $2 \times 10^{-4}$, except for segmentors which adopt a decay rate of 0.9 for every two epochs.
The hyperparameters $\lambda_{cyc}$, $\lambda_{kd}^{1}$ and $\lambda_{kd}^{2}$ are empirically set as $10$, $0.5$ and $1$, respectively.

\subsection{Experimental Results}
\subsubsection{Comparison with other methods.}
To demonstrate the effectiveness of our method, we train a segmentor with only CT data as our baseline and compare our method with other multi-modality learning methods.
The quantitative results in whole heart segmentation are shown in Table~\ref{tab:quatitative}.
In joint-training, the segmentor is trained with both CT and MRI data simultaneously, while in fine-tune, we pretrain a segmentor with MRI data and then update the network parameters by CT data.
We compare with X-shape architecture, which achieves the best segmentation performance among all dual-stream architectures in~\cite{valindria2018multi}.
We also compare with~\citeauthor{jiang2018tumor}~\shortcite{jiang2018tumor} and Zhang et al.~\shortcite{zhang2018translating}, which employ GAN to alleviate multi-modality appearance discrepancy in network training.
As $S_{syn}$ and $S_{real}$ acquire two modality prior knowledge, we input CT data in both segmentors. Considering complementary information, we adopt the ensemble prediction as final results.
Note that all methods mentioned above are experimented with the same segmentor backbone for fair comparison.
We use the Dice coefficient as evaluation metric.

It is observed that with assisted modality data, all multi-modality learning approaches achieve average dice improvements of all substructures in CT segmentation compared with baseline, which demonstrates the feasibility of utilizing assisted modality priors to promote the performance on target-modality segmentation.
However, the mean dice improvements of fine-tune and joint-training are limited, \ie, $0.63\%$ and $0.37\%$, respectively. X-shape method only achieves comparable results with fine-tune, since separate branch could only avoid, not directly reduce appearance gap.
Although \citeauthor{jiang2018tumor}~\shortcite{jiang2018tumor} proposed to use GAN for multi-modality appearance difference reduction, the improvement is still limited. The reason may be that they train the GAN and segmentor in offline manner, where the segmentor would be incompatible to the generated images.
With the help of online training strategy, Zhang et al.~\shortcite{zhang2018translating} achieve $0.85\%$ mean dice improvements than \citeauthor{jiang2018tumor}~\shortcite{jiang2018tumor}.
Compared with Zhang et al.~\shortcite{zhang2018translating}, our approach takes advantages of mutual knowledge distillation scheme and can interactively improves the cross-modality knowledge exploration of each segmentor
Therefore, our method (last row in Table~\ref{tab:quatitative}) achieves the highest dice in all heart substructure segmentation and further outperforms Zhang et al.~\shortcite{zhang2018translating} with $1.62\%$ in mean dice.

For our approach, both real segmentor $S_{real}$ and synthetic segmentor $S_{syn}$ enhance the mean dice with $2.41\%$ and $1.43\%$ compared to baseline.
Since $S_{real}$ is trained by direct supervision from real CT annotations, while $S_{syn}$ directly learn from synthetic CT data, the mean dice of $S_{real}$ is slightly better than $S_{syn}$.
During training, $S_{real}$ learns explicit knowledge from CT data and implicit priors from MRI data. Meanwhile, $S_{syn}$ absorbs explicit MRI knowledge and implicit CT priors. Considering the complimentary information, the ensemble segmentor acquires $3.06\%$ improvements than baseline in mean dice, which demonstrates the effectiveness of our method.
One intuitive explanation of why our mutual knowledge distillation performs better than joint-training is that joint training could be seen as ``hard-parameter" sharing (same parameters) between real and synthetic segmentor, while our Mutual Knowledge Distillation scheme enables ``soft-parameter" sharing, which provides more flexibility in feature extraction.

It is worth to mention that the effectiveness of our method is from the cross-modality mutual knowledge distillation, not solely gains from model ensemble. For single segmentor, our $S_{real}$ outperforms Zhang et al.~\shortcite{zhang2018translating} with $1.27\%$ in mean dice with the same segmentor architecture, as shown in Table~\ref{tab:quatitative}. Moreover, for ensemble segmentor, we train two segmentors in Zhang et al.~\shortcite{zhang2018translating} with the combination of synthetic and real CT and use the ensemble prediction as the final results. This ensemble method achieve the mean dice of $88.92\%$, inferior to our final results by $1.2\%$, validating the effectiveness of proposed MKD regularization scheme.
\begin{table}[tbp]
    \centering
    \caption{Ablation study of key components in our method, where the mean dice of all heart substructures is reported.}\smallskip
    \label{tab:Ablation}
    \begin{tabular}{p{2.5cm}<{\centering}|c|c|c}
    \toprule[1pt]
    Methods & $S_{syn}$ & $S_{real}$ & $S_{ens}$ \\
    \midrule
    W/o IAM & $0.8730$ & $0.8745$ & $0.8829$ \\
    W/o $\mathcal{L}_{kd}^{s \rightarrow r}$ & $0.8825$ & $0.8742$ & $0.8879$ \\
    W/o $\mathcal{L}_{kd}^{r \rightarrow s}$ & $0.5785$ & $0.8904$ & $0.7937$\\
    Ours & $\bm{0.8849}$ & $\bm{0.8947}$ & $\bm{0.9012}$ \\
    \bottomrule[1pt]
    \end{tabular}
\end{table}

\begin{table}[t]
    \centering
    \caption{Performance comparison of different amounts of assisted modality data in training.}\smallskip
    \label{tab:Extensive}
    \begin{tabular}{p{2cm}<{\centering}|p{2.5cm}<{\centering}}
    \toprule[1pt]
    Setting & Mean Dice \\
    \midrule
    5-MRI & $0.8916$ \\
    10-MRI & $0.8945$ \\
    20-MRI & $\bm{0.9012}$ \\
    \bottomrule[1pt]
    \end{tabular}
    \vspace{-0.5cm}
\end{table}

\subsubsection{Qualitative comparison.} The qualitative comparison between different methods is presented in Figure~\ref{fig:qualitative}.
As observed, without assistant-modality prior knowledge, the network trained only with CT data would misidentify many substructures.
After exploiting assisted modality information, the segmentation results are better but still scattered and fragmentary, especially in fine-tune and X-shape.
Owing to the proposed Mutual Knowledge Distillation scheme, the predictions of our method have more smooth contours and are closer to the ground truth compared to others.
\subsubsection{Ablation analysis.}
\begin{figure}[htbp]
\centering
\includegraphics[width=0.48\textwidth]{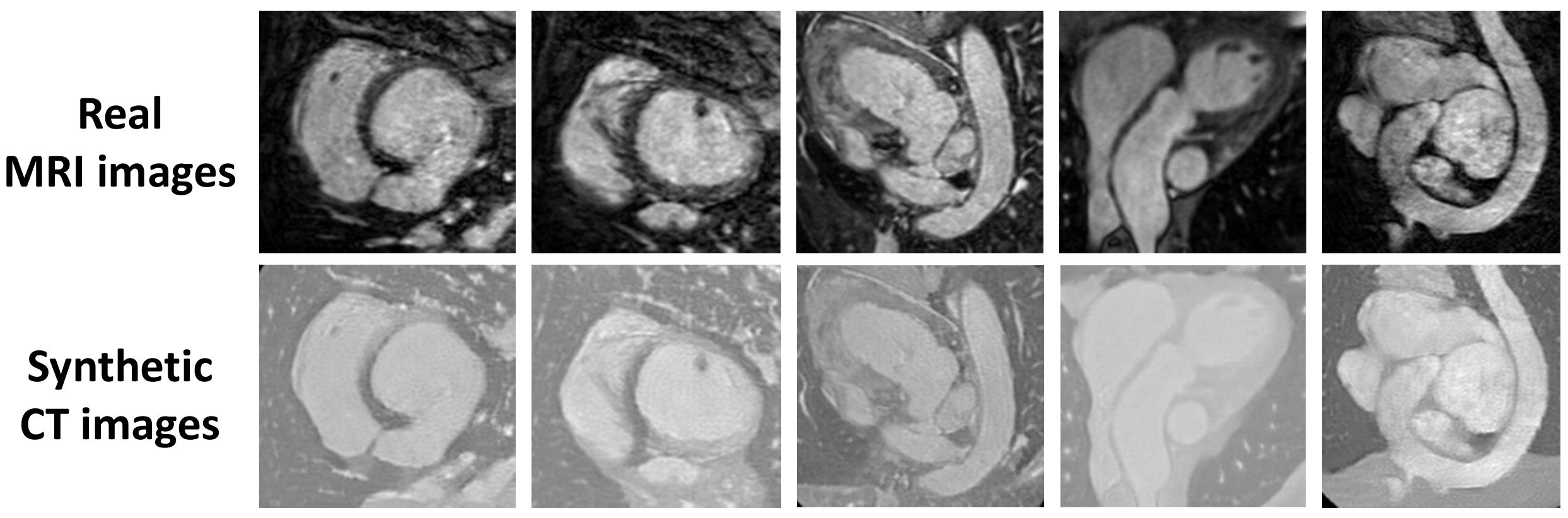}
\caption{Generated results of Image Alignment Module, where MRI structures in synthetic CT images are well-preserved with clear edges.}
\label{fig:appearance}
\end{figure}

To investigate the effectiveness of the key components in our framework, we conduct ablation studies and the quantitative results are presented in Table~\ref{tab:Ablation}.
We first remove the Image Alignment Module (IAM) and directly train $S_{syn}$ and $S_{real}$ with raw CT and MRI data under the Mutual Knowledge Distillation scheme.
As observed from Table~\ref{tab:Ablation}, the mean dice of $S_{syn}$ and $S_{real}$ are similar to each other. The segmentation performance of $S_{syn}$, $S_{real}$ and $S_{ens}$ are degraded by $1.19\%$, $2.02\%$ and $1.83\%$, respectively, showing the importance of IAM in our framework.
We further present some synthesized CT images in Figure~\ref{fig:appearance}. As we can see, the heart structures are clear and well-preserved during translation.
In addition, we also conduct experiments with only one-way knowledge distillation existing in our framework.
As shown in Table~\ref{tab:Ablation}, without synthetic-to-real knowledge distillation, the results of $S_{syn}$ is comparable with mutual distillation, while the performance of $S_{real}$ is weaken and compromised. Specifically, $S_{real}$ and $S_{ens}$ decrease $2.05\%$ and $1.33\%$ compared to mutual distillation respectively.
In another aspect, with the absence of real-to-synthetic knowledge distillation, the performance of $S_{syn}$ is hugely deteriorated and further corrupts $S_{ens}$ performance.
As no real CT data is involved in any ways for training $S_{syn}$, the $S_{syn}$ is completely isolated and lost.
With our proposed mutual knowledge distillation, both segmentors could gain mutual benefits from each other's training.

\begin{figure}[ht]
\centering
\includegraphics[width=0.48\textwidth]{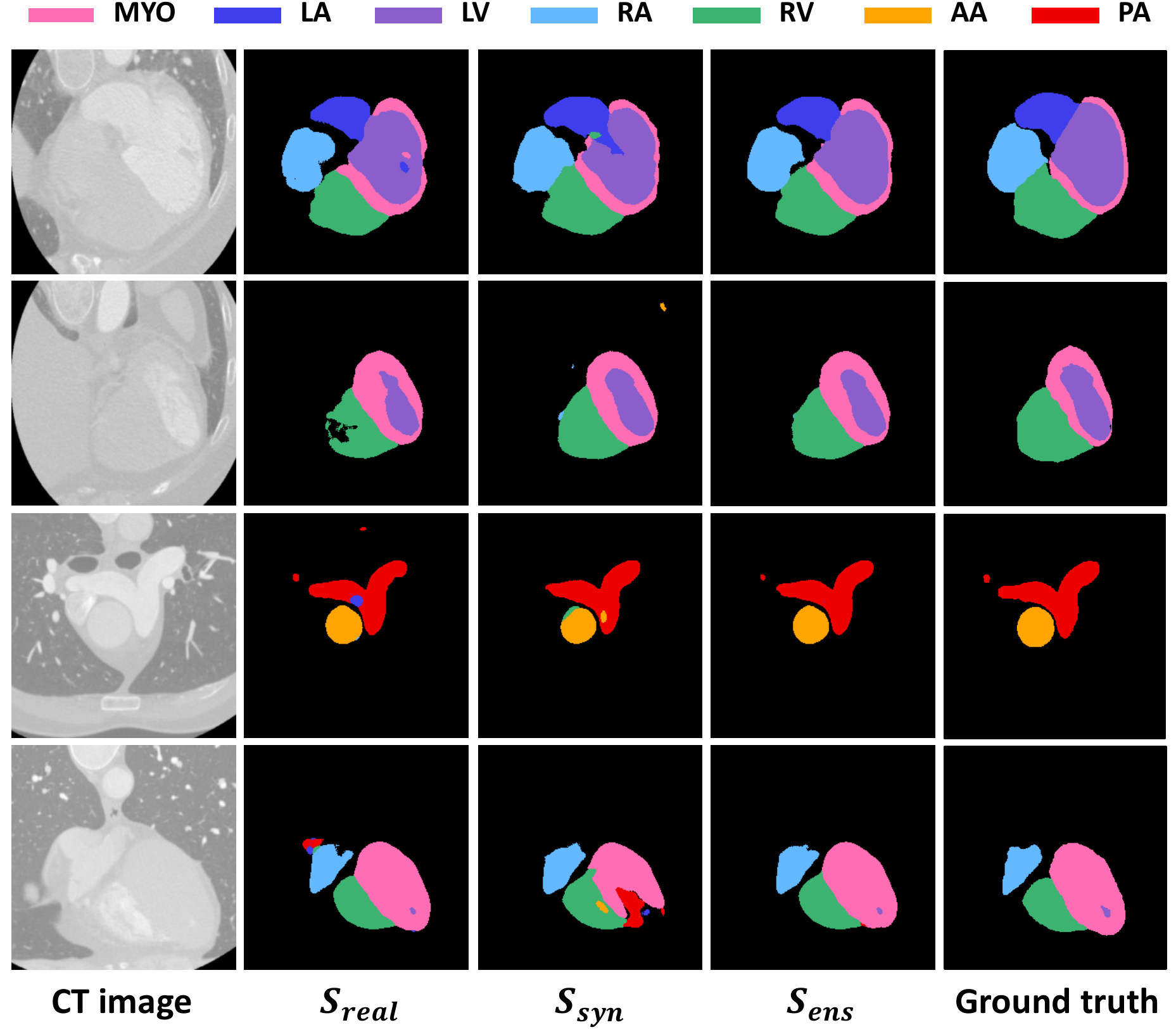}
\caption{Visual comparison of segmentation results by $S_{real}$, $S_{syn}$ and the ensemble segmentor, which could balance inconsistent predictions and produce reliable results.}
\label{fig:ablation}
\end{figure}

We also present the qualitative comparison among predictions of $S_{syn}$, $S_{real}$ and our ensemble results in Figure~\ref{fig:ablation}.
Since the cohort of $S_{syn}$ and $S_{real}$ integrates information from annotations and peer guidance from each other's outputs,
the ensemble segmentor could highlight the uniform predictions while correcting the misleading predictions.
For those inconsistent predictions of $S_{syn}$ and $S_{real}$ (\eg, last row in Figure~\ref{fig:ablation}), $S_{ens}$ comprehensively considers different opinions and gives back reliable predictions.
Overall, the key components of our network are tightly incorporated and collaboratively devote to remarkable segmentation results.

\begin{table*}[htp]
\centering
\caption{Quantitative comparison between our approach and other multi-modality segmentation methods in whole heart segmentation. Here we take MR as target modality and CT as assistant modality. The dice of all cardiac substructures and mean dice are reported.}
\smallskip
\label{tab:quatitative_CT_to_MR}
\begin{tabular}{c|c|c|c|c|c|c|c|c}
\toprule[1pt]
\multirow{2}{*}{Method} & \multirow{2}{*}{Mean Dice} & \multicolumn{7}{c}{Dice of substructures of heart} \\ \cline{3-9} 
 & & MYO & LA & LV & RA & RV & AA & PA\\
\midrule
Baseline & $0.8243$ & $0.8259$ & $0.8153$ & $0.9347$ & $0.8146$ & $0.9140$ & $0.7365$ & $0.7288$\\
Fine-tune & $0.8217$ & $0.8336$ & $0.8447$ & $0.9371$ & $0.7893$ & $0.8940$ & $0.7015$ & $0.7516$\\
Joint-training & $0.8306$ & $0.8375$ & $0.7893$ & $0.9303$ & $0.8507$ & $0.9119$ & $0.7487$ & $0.7455$\\
X-shape ~\shortcite{valindria2018multi} & $0.8337$ & $0.8321$ & $0.8247$ & $0.9347$ & $0.8450$ & $0.9181$ & $0.7278$ & $0.7536$\\
Jiang et al.~\shortcite{jiang2018tumor} & $0.8370$ & $0.8309$ & $0.8569$ & $0.9356$ & $0.8582$ & $0.9016$ & $0.7334$ & $0.7422$\\
Zhang et al.~\shortcite{zhang2018translating} & $0.8375$ & $0.8360$ & $0.8169$ & $0.9317$ & $0.8501$ & \bm{$0.9183$} & $0.7435$ & \bm{$0.7663$} \\
\midrule
Ours ($S_{syn}$) &  $0.8382$ & $0.8405$ & $0.8308$ & $0.9337$ & \bm{$0.8655$} & $0.9136$ & $0.7351$ & $0.7483$ \\
Ours ($S_{real}$) & $0.8466$ & $0.8507$ & $0.8582$ & $0.9427$ & $0.8623$ & $0.9091$ & $0.7398$ & $0.7633$ \\
Ours & \bm{$0.8517$} & \bm{$0.8525$} & \bm{$0.8695$} & \bm{$0.9437$} & \bm{$0.8655$} & $0.9121$ & \bm{$0.7544$} & $0.7640$\\
\bottomrule[1pt]
\end{tabular}
\end{table*}

\subsubsection{Effect of different amounts of assistant-modality data.}
We further investigate the effects of different amounts of assistant-modality data in the network training.
We conduct experiments with 5, 10 and 20 MRI volumes as assistant-modality data in network training and quantitatively evaluate these settings on CT segmentation in Table~\ref{tab:Extensive}.
Like the previous, we use the same CT volumes and perform two-fold cross-validation, and randomly pick the corresponding number of MRI volumes as assisted modality data.
Generally, it is observed that the performance of CT segmentation is enhanced with the increase of MRI data, as more MRI data implies more assisted prior knowledge.

\subsubsection{Extensive experiments with CT as assistant modality.}
To demonstrate the benefit of our proposed mutual knowledge distillation is bidirectional, we conduct another experiments to take CT as assistant modality, while MR as target modality. All substructure dice and their mean dice are reported in Table~\ref{tab:quatitative_CT_to_MR}. Our method outperforms baseline and Zhang et al.~\shortcite{zhang2018translating} with $2.74\%$ and $1.42\%$ in mean dice, respectively.
Our single segmentor $S_{real}$ also achieves better performance than Zhang et al.~\shortcite{zhang2018translating} by $0.91\%$ in average dice.
The results show that our method could be bidirectionally applied to both modalities and have no constrain in target modality chosen.

\section{Conclusion}
In this paper, we propose a novel multi-modality image segmentation framework to demonstrate the effectiveness of exploiting assistant-modality prior knowledge to enhance the segmentation performance on target modality.
In specific, we first design an Image Alignment Module to reduce appearance discrepancy by generating synthetic target-modality images via adversarial learning.
More importantly, we present the Mutual Knowledge Distillation scheme to thoroughly exploiting the modality-shared knowledge by explicitly learning one modality knowledge from segmentation annotations and implicitly exploring another modality knowledge with peer guidance.
The experimental results on public multi-modality cardiac segmentation dataset verify the effectiveness of our method and demonstrate superior performance against other state-of-the-art methods.

\section{Acknowledgements}
This work described in this paper was supported by the following
grants from the Hong Kong Innovation and Technology Commission
(Project No. ITS/426/17FP) and the National Natural Science Foundation of China (Project No. U1813204).


\bibliographystyle{aaai}
\bibliography{ref}

\end{document}